\documentclass[journal]{IEEEtran}
\usepackage{cite}
\usepackage{mathptmx}
\usepackage[dvips]{graphicx}
\usepackage{url}

\newcommand{\Ref}[1]{Ref.~\citen{#1}}
\newcommand{\Fig}[1]{Fig.~\ref{#1}}

\newcommand{\Tab}[1]{Table~\ref{#1}}

\newcommand{\EA}[5]{\ensuremath{
\mbox{#1}^{+\mbox{\scriptsize{#2}}}_{-\mbox{\scriptsize{#3}}}
\times\mbox{10}^{#4\mbox{\scriptsize{#5}}}}}

\begin{document}

\title{The Large-Angle Photon Veto System for the NA62 Experiment at CERN}
\urldef{\moulson}\url{moulson@lnf.infn.it}
\author{F.~Ambrosino, B.~Angelucci, A.~Antonelli, F.~Costantini, G.~D'Agostini,
D.~Di~Filippo, R.~Fantechi, S.~Gallorini, S.~Giudici, E.~Leonardi, 
I.~Mannelli, P.~Massarotti, M.~Moulson$^*$, M.~Napolitano, V.~Palladino, 
F.~Rafaelli, M.~Raggi, G.~Saracino, M.~Serra, T.~Spadaro, P.~Valente, 
S.~Venditti
\thanks{Manuscript received November 15, 2011.}%
\thanks{F.~Ambrosino, D.~Di~Filippo, P.~Massarotti, M.~Napolitano, and 
G.~Saracino are with the Dipartimento di Scienze Fisiche dell'Universit\`a 
and Sezione INFN, Napoli, Italy.}%
\thanks{B.~Angelucci, F.~Costantini, R.~Fantechi, S.~Gallorini, S.~Giudici, 
I.~Mannelli, F.~Rafaelli, and S.~Venditti are with the Dipartimento di Fisica
dell'Universit\`a and Sezione INFN, Pisa, Italy.}%
\thanks{A.~Antonelli, M.~Moulson, M.~Raggi, and T.~Spadaro are with the 
Laboratori Nazionali di Frascati dell'INFN, Frascati, Italy.}%
\thanks{G.~D'Agostini, E.~Leonardi, V.~Palladino, M.~Serra, and P.~Valente 
are with the Dipartimento di Fisica dell'Universit\`a ``La Sapienza'' 
and Sezione INFN, Roma, Italy.}%
\thanks{$^*$\,Presenter. Address correspondence to Matthew Moulson, e-mail: 
\moulson}
}

\maketitle
\thispagestyle{empty}

\begin{abstract}
The branching ratio (BR) for the decay $K^+\to\pi^+\nu\bar{\nu}$ is a 
sensitive probe for new physics. The NA62 experiment at the CERN SPS will 
measure this BR to within about 10\%. To reject the dominant background 
from channels with final state photons, the large-angle vetoes (LAVs) must 
detect photons of energy as low as 200 MeV with an inefficiency of 
less than 10$^{-4}$, as well as provide energy and time measurements 
with resolutions of 10\% and 1~ns for 1 GeV photons. The LAV 
detectors make creative reuse of lead glass blocks recycled from the OPAL 
electromagnetic calorimeter barrel. We describe the mechanical design 
and challenges faced during construction, the characterization of the 
lead glass blocks and solutions adopted for monitoring their performance, 
and the development of front-end electronics to allow simultaneous time 
and energy measurements over an extended dynamic range using the 
time-over-threshold technique. Our results are based on test-beam data
and are reproduced by a detailed Monte Carlo simulation that includes 
the readout chain.
\end{abstract}

\begin{IEEEkeywords}
Calorimetry, Detectors, Elementary particles
\end{IEEEkeywords}

\section{The NA62 Experiment}

The decays $K^+\to\pi^+\nu\bar{\nu}$ and $K_L\to\pi^0\nu\bar{\nu}$ are
flavor-changing neutral-current processes whose amplitudes are dominated 
by $Z$-penguin and box diagrams. Because there are no contributions from 
long-distance processes with intermediate photons and because the hadronic 
matrix elements can be obtained from rate and form factor measurements 
for common $K\to\pi\ell\nu_\ell$ decays, the branching ratios (BRs) for the
$K\to\pi\nu\bar{\nu}$ decays can be calculated in the Standard Model (SM) 
with minimal intrinsic uncertainty 
(see \Ref{C+11:kaonRev} for a recent review). 
The BRs for these decays are therefore a sensitive probe of the SM flavor 
sector and provide constraints on the CKM unitarity triangle that are 
complementary to those from measurements of $B$-meson decays. 

\begin{figure}
\centering
\includegraphics[width=0.9\linewidth]{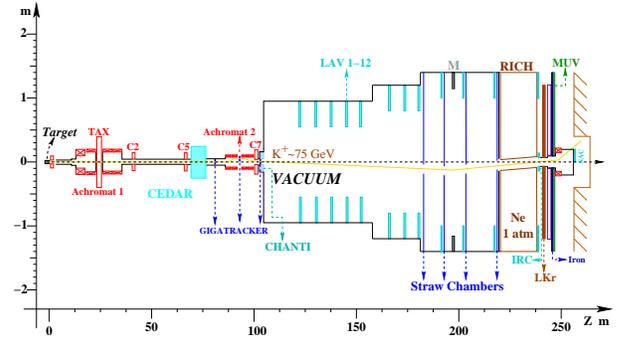}
\caption{The NA62 experimental layout.}
\label{fig:expt}
\end{figure}

On the other hand, the tiny BRs for these decays are notoriously difficult
to measure, not least because of the three-body kinematics with two
undetectable neutrinos in the final state. At present, the experimental value 
of the BR for the decay 
$K^+\to\pi^+\nu\bar{\nu}$ is \EA{1.73}{1.15}{1.05}{-}{10} on the basis 
of seven detected candidate events \cite{E949+08:Kpnn2}.
The goal of NA62, an experiment at the CERN SPS,
is to detect $\sim$100 $K^+\to\pi^+\nu\bar{\nu}$ decays
with a S/B ratio of 10:1 
in two years of data taking beginning in 2013.
The experiment is fully described in \Ref{NA62+10:TDD}.
The experimental layout is illustrated in \Fig{fig:expt}.

NA62 will make use of a 75-GeV unseparated positive secondary beam
with a total rate of nearly 800~MHz, of which $\sim$50 MHz is $K^+$'s.
The $K^+$'s are identified by the CEDAR differential Cerenkov counter in the
beamline. All 800~MHz of beam particles are tracked by three silicon-pixel 
tracking detectors (the Gigatracker) located at the achromat just upstream 
of the vacuum decay volume, providing event-by-event measurements of the 
$K^+$ trajectory and momentum. The decay volume is evacuated to $10^{-6}$ 
mbar in order to minimize background between interactions and residual gases.
It begins $\sim$100~m downstream of the production target, is $\sim$110~m 
long, and consists of segments of increasing diameter, from $\sim$2~m
upstream to $\sim$3~m downstream.
5~MHz of kaon decays are observed in the 65-m fiducial decay region, in 
the upstream part of the vacuum tank. The ring-shaped large-angle photon 
vetoes (LAVs) 
are placed at 11 stations along the vacuum volume and provide full coverage 
for decay photons with $8.5~{\rm mrad}<\theta<50~{\rm mrad}$.
The last 35~m of the vacuum volume host a dipole
spectrometer with four straw-tracker stations operated in the vacuum.
At the exit of the vacuum region, a ring-imaging Cerenkov counter (RICH) 
4~m in diameter by 17~m in length helps to identify charged decay secondaries.
Downstream of the RICH, a number of photon vetoes provide hermeticity,
including principally the NA48 liquid-krypton calorimeter (LKr) to veto 
forward ($1~{\rm mrad}<\theta<8.5~{\rm mrad}$), high-energy photons.
The 12th LAV station provides downstream coverage at large angles 
($8.5~{\rm mrad}<\theta<50~{\rm mrad}$), while a ring-shaped shashlyk 
calorimeter (IRC) about the 
beamline provides coverage for photons with $\theta<1~{\rm mrad}$.
Further downstream, a muon veto detector (MUV) provides additional rejection
for $K\to\mu\nu$ events, and a small-angle shashlyk calorimeter (SAC) 
around which the beam is deflected completes the coverage for 
very-small-angle photons that would otherwise escape via the beam pipe.

Assuming an acceptance for signal events of about 10\%, the experiment 
must be able to reject background from the dominant $K^+$ decays such as
$K^+\to\pi^+\pi^0$ at the level of $10^{12}$.
Kinematic cuts on the $K^+$ and $\pi^+$ tracks (as reconstructed in the
Gigatracker and straw chambers, respectively) provide a rejection 
factor of $10^4$ and ensure that the photons from the $\pi^0$ have 40~GeV
of energy. There is a kinematic correlation between photon energy and angle of
emission with respect to the beam axis; the forward photons that are 
intercepted by the LKr calorimeter, IRC, and SAC have much higher energies than
those intercepted by the LAVs. Nevertheless, the photons from 
$K^+ \to \pi^+\pi^0$ intercepted by the LAVs may have energies from a 
few tens of MeV to several GeV. 
In order to detect the $\pi^0$ with an inefficiency of $\leq 10^{-8}$,
the maximum tolerable inefficiency in the LAV detectors for photons 
with energies as low as 200~MeV is $10^{-4}$.
In addition, the LAV detectors must have good time resolution ($\sim$1 ns)
to allow signals from incident particles to be identified with the
correct event, and good energy resolution ($\sim$10\% at 1 GeV) for precise
vetoing and use in full-event reconstruction. 
The system must also be sensitive to minimum-ionizing particles.
Finally, the LAV detectors must be compatible with operation in a vacuum 
of $10^{-6}$ mbar.

\section{The Large-Angle Veto System}

The NA62 LAV detectors make creative
reuse of lead glass blocks recycled from the OPAL electromagnetic calorimeter
barrel\cite{OPAL+91:NIM}, which became available in 2007, when various 
technologies were under consideration for the construction of the LAV 
detectors. Other solutions considered included a lead/scintillating tile design 
originally proposed for use in the (later canceled) CKM experiment at Fermilab, 
and a lead/scintillating-fiber design based on the electromagnetic calorimeter
for the KLOE experiment. Prototype instruments based on each 
of the three technologies were obtained or constructed, and tested
with the electron beam at the Frascati Beam-Test Facility. 
These tests demonstrated that all three technologies are suitable 
for use in NA62 \cite{A+07:Veto}. In particular, the inefficiency for the 
detection of single, tagged electrons with the OPAL lead glass modules
was measured to be \EA{1.2}{0.9}{0.8}{-}{4} at 203 MeV and
\EA{1.1}{1.9}{0.7}{-}{5} at 483 MeV.
Basing the construction of the LAV system on the OPAL lead glass modules 
provides significant economic advantages.

\begin{figure}
\centering
\includegraphics[width=0.85\linewidth]{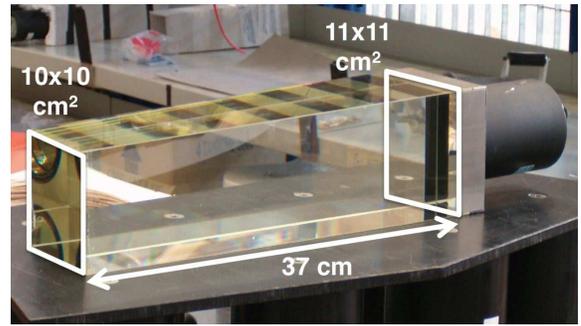}
\caption{A module from the OPAL calorimeter, without wrapping and with 
reinforcement plates at the interface between the glass and the steel 
flange.}
\label{fig:block}
\end{figure}
The modules from the central part of the OPAL 
electromagnetic calorimeter barrel consist of blocks of Schott SF57 lead glass.
This material is about 75\% lead oxide by weight and has a density
$\rho = 5.5~{\rm g}/{\rm cm}^3$ and a radiation length $X_0 = 1.50$~cm;
its index of refraction is $n \approx 1.85$ at $\lambda = 550~{\rm nm}$
and $n \approx 1.91$ at $\lambda = 400~{\rm nm}$.
Electromagnetic showers in the lead glass are detected by virtue 
of the Cerenkov light produced; our measurements indicate that, averaged over
modules, minimum ionizing particles produce about 0.34~p.e. per MeV of 
deposited energy. 
The front and rear faces of the blocks measure about 
$10\times10$ cm$^2$ and
$11\times11$ cm$^2$, respectively; the blocks are 37~cm in length. (The precise
geometry depends slightly on the ring of the OPAL calorimeter from which
each block is extracted; blocks of uniform geometry are used in the 
construction of each ring of the LAV system.)
Each block is read out at the back side by a Hamamatsu R2238 76-mm 
PMT, which is optically coupled via a 4-cm long cylindrical light 
guide of SF57 of the same diameter as the PMT.
The rear face of the glass block is glued to a 1-cm thick stainless 
steel flange featuring the holes that allow mounting 
hardware to be attached and a circular cutout for the light guide. 
A mu-metal shield surrounding the PMT and light guide is also glued to the
flange. A complete module (block plus PMT) is a monolithic assembly; the 
block and PMT cannot be independently replaced. 
Figure~\ref{fig:block} shows a picture of a complete module.

\begin{figure}
\centering
\includegraphics[width=0.45\linewidth]{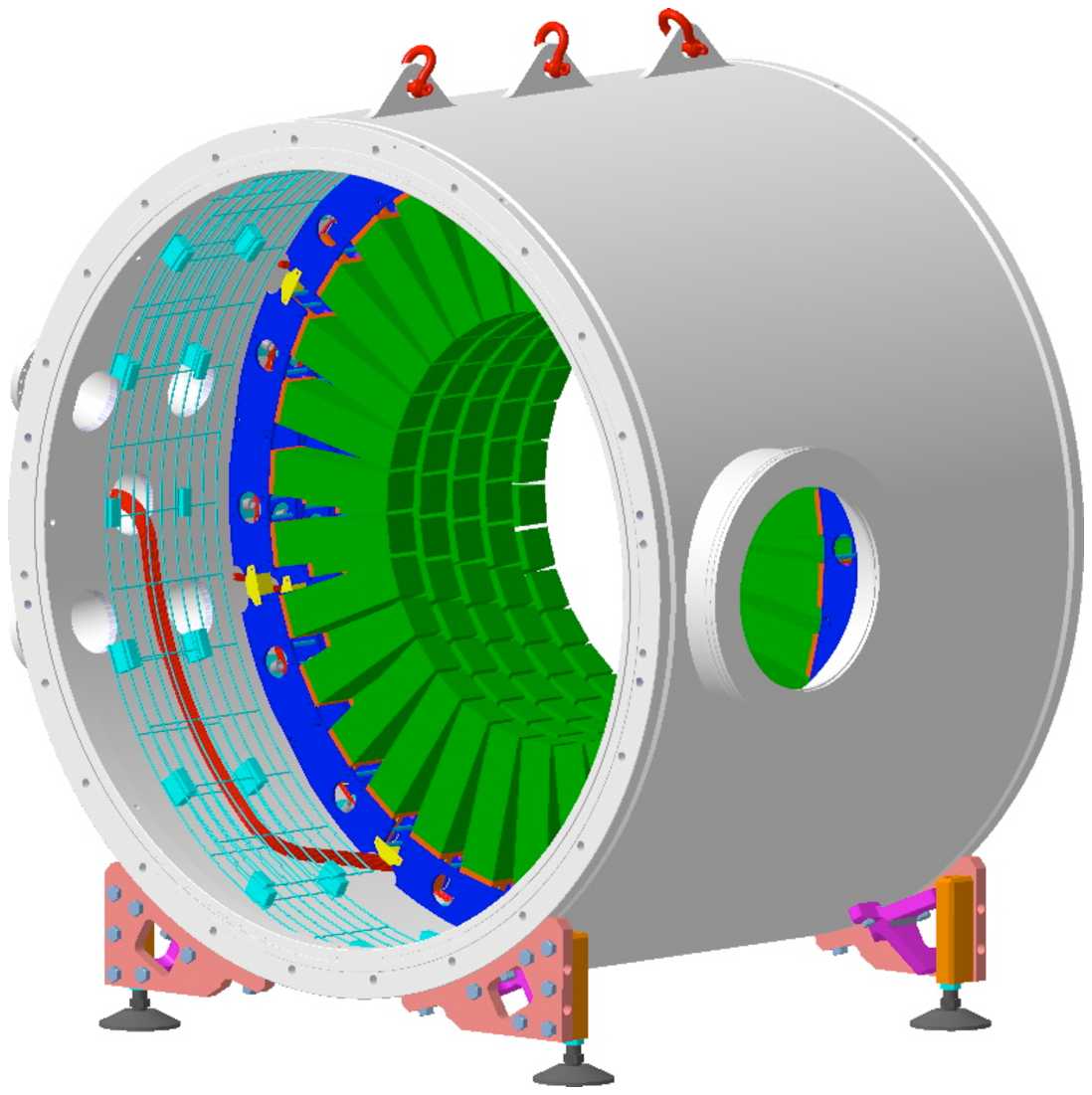}
\hspace{0.10\linewidth}
\includegraphics[width=0.39\linewidth]{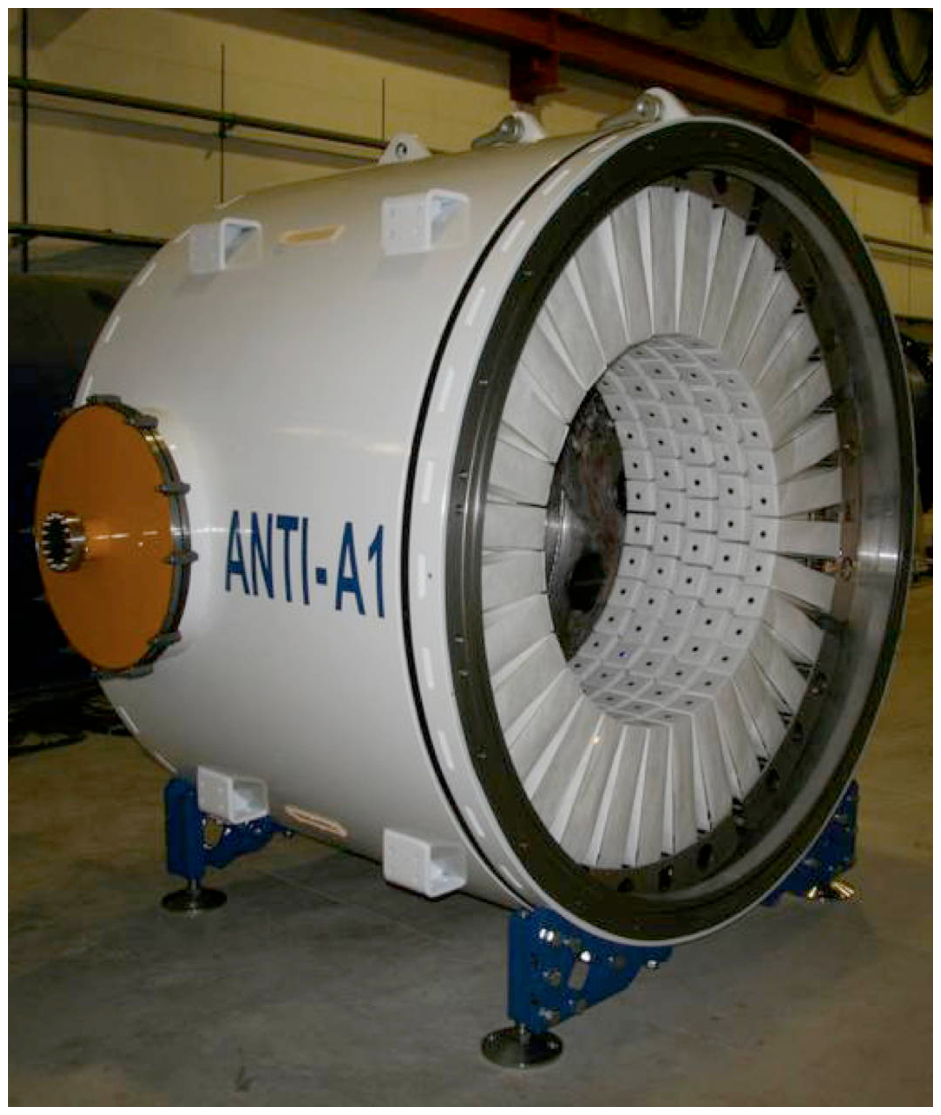}
\caption{Design study (left) and completed prototype A1 veto station (right)
making use of the OPAL lead glass calorimeter elements.}
\label{fig:station}
\end{figure}
A LAV station is made by arranging these blocks around the inside of 
a segment of vacuum tank, with the blocks aligned radially to form an 
inward-facing ring. Multiple rings are used in each station in order to
provide the desired depth for incident particles. The blocks in successive
rings are staggered in azimuth; the rings are spaced longitudinally by 
about 1 cm.

\begin{table}
\renewcommand{\arraystretch}{1.3}
\caption{Parameters of LAV Stations}
\label{tab:param}
\centering
\begin{tabular}{@{}llllll@{}}
\hline\hline
Station & Diameter [mm] & \multicolumn{2}{l}{Block radius [mm]} & Layers & Blocks\\
 & Outer wall & Inner & Outer & & \\
\hline
A1--A5  & \phantom{$\sim$}2168  & \phantom{1}537 & \phantom{1}907 & 5 & 160\\
A6--A8  & \phantom{$\sim$}2662  & \phantom{1}767 &           1137 & 5 & 240\\
A9--A11 & \phantom{$\sim$}3060  & \phantom{1}980 &           1350 & 4 & 240\\
A12     &           $\sim$3250  &           1072 &           1442 & 4 & 256\\
\hline\hline
\end{tabular}
\end{table}
The LAV system consists of a total of 12 stations, the diameter of which 
increases with distance from the target.
The geometry of the LAV stations is summarized in \Tab{tab:param}.
For stations A1--A5, A6--A8, and A9--A11, apart from the different sizes and 
block configurations, the designs are conceptually similar.
The geometry of A12 is not yet final; this station is operated in air and
its design is different from that of the other stations. 
Since the spaces between the blocks are significantly smaller in the 
larger-diameter vessels, fewer layers are necessary. As a result of the 
staggering scheme, particles incident on any station are intercepted by
blocks in at least three rings, for a total minimum effective depth of 
21 radiation lengths. The vast majority of incident particles are intercepted
by four or more blocks (27\,$X_0$).
The stations with five layers (A1--A8) are 1.55~m in length, while  
those with four layers (A9--11) are 1.43~m in length.  

Station A1 was constructed as a prototype during the first half of 2009.
It was installed in the NA62 beamline at CERN and tested with electrons 
and muons in October 2009.
The design study and completed station are shown in \Fig{fig:station}.
On the basis of experience gained during the test beam, various 
improvements were made. Station A2 was then constructed and tested with
an unseparated, low-energy positive beam in the T9 beamline at the CERN PS in 
August 2010. A1 was subsequently rebuilt to incorporate the design 
improvements. Construction of the remaining stations was then commenced and
is now underway.

\section{LAV Construction}

The LAV vacuum vessels are constructed from 25-mm thick steel plate, 
rolled into a cylinder and arc welded along the seam. Structural steel 
(S275JR) is used for stations A1 to A8 and A11. Stations A9 and A10 are 
mounted astride the spectrometer magnet, so non-magnetic 
($\mu/\mu_0 \approx 1.01$) 304L stainless steel is used for these stations.
After rolling, seam welding, and welding of the end flanges, the interior is 
turned on a vertical lathe and the vessel is vacuum tested. Five 
200-mm ISO-F flanges for HV, signal, and calibration connections are 
welded via nozzles 20-cm long to one side of the vessel. One 630-mm ISO-K flange
is attached to the opposite side for use as a manhole; the pumps for the 
NA62 vacuum system are also mounted on these flanges. Bolt holes on the 
flanges and in the interior walls of the tank are drilled with
a CNC milling machine. A final vacuum test is performed, the residual
magnetic field is mapped, and the chamber is cleaned and the outside is 
painted. A passivator that does not interfere with high-vacuum operation is 
applied to the inside walls of the vessel. The contract for the construction 
of vessels A1 to A8 and A11 was awarded to Fantini SpA (Anagni, Italy).
Construction of A1 to A7 has been completed; construction of A8 is in progress. 

The OPAL detector modules (lead glass block plus PMT) were manufactured 
by Hamamatsu during the mid-1980s. Their recycling requires substantial 
care throughout the assembly procedure.

After the storage area in which the modules were kept was inundated 
during a flash flood in spring 2008, the modules were subject to an 
extensive sorting and clean-up effort carried out at CERN by an industrial 
recovery firm. They were subsequently shipped to Frascati for use in LAV
assembly.

The interface between the stainless steel flange and lead glass block is 
fragile, and is found to be critically damaged in a few percent of the modules 
upon first examination. Typically, the epoxy between the glass and steel is 
found to have become delaminated in some areas, while in others, the glass is 
fractured at the interface with pieces still adhering to the flange. This 
is attributed to thermally induced stress from the differing expansion
coefficients of the steel and of the glass and/or thermal shock in the glass
due to the conductivity of the flange. Some modules are found to have 
the glass block completely separated from the steel flange. Since 
the modules are suspended by the flange in the LAV system, the fragility of the
interface is a structural vulnerability and a safety risk.
The first step in the processing of the modules at Frascati is to therefore
to reinforce the interface. Using epoxy resin, 20~cm$^2 \times 0.5$-mm thick 
stainless steel plates are attached across the glass-steel interface 
on all four sides of the block. Calculations indicate and static tests 
confirm that the reinforced bond is several times stronger than the original
bond.

To maintain the design vacuum of $\sim$10$^{-6}$~mbar, all components of the
LAV detectors must be carefully cleaned. A variety of techniques are used, 
including pickling, solvent degreasing, and ultrasonic cleaning. 
The lead glass blocks require special care. After reinforcement of the 
glass/steel interface, the original wrapping is removed and the blocks 
are thoroughly cleaned with acetone or isopropyl alcohol. They are then 
wrapped with a new laser-cut and heat-welded Tyvek cover.

During the A1 test beam, ringing of the analog signal was observed to lead 
to errors in charge reconstruction using the time-over-threshold technique
discussed below. This problem was traced to a small parasitic inductance in 
the PMT dynodes and solved by replacing the original OPAL HV dividers soldered 
to the PMTs with new dividers of our own design. The new divider features 
additional resistors on the last three dynodes and anode to damp out the 
oscillation, storage capacitors for the last three dynode stages to improve 
response linearity for large signals, and a decoupling resistor between the 
HV and signal grounds to decrease noise. 

\begin{figure}
\centering
\includegraphics[width=0.725\linewidth]{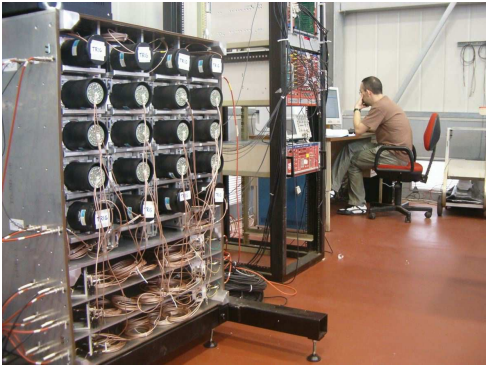}
\caption{Test stand used for the characterization of lead glass detector 
modules. The top and bottom rows of lead glass blocks form a telescope for 
the selection of vertical cosmic rays. Here, the front panel is open; it 
is closed during operation.}
\label{fig:oven}
\end{figure}
After the divider is replaced, the blocks are tested and characterized 
12 at a time using a test station featuring an LED pulser and cosmic-ray 
telescope (\Fig{fig:oven}).
The PMT gains are measured first, by varying the intensity of the light 
pulses from the LEDs as well as the PMT HV settings and mapping out the 
response for each block. Using the gain curves so obtained, the PMTs
are then set to a reference value of the gain ($9\times10^5$ or $1\times10^6$)
and the response to cosmic rays selected by the telescope is measured; the
photoelectron yield for the block (p.e./MeV) is then obtained assuming that
vertically incident cosmic rays leave 77 MeV in each block. As noted above,
photoelectron yields of 0.34~p.e./MeV are typical.
Finally, using the gain curves 
and the measurements of photoelectron yield, the PMT voltages are 
set to the values expected to produce a common output charge level of 4.5 pC
for cosmic ray events. The response is measured and the HV setting is 
validated. Thus, at the end of a 12-hour cycle, which is fully automated 
using LabView, we have PMT gain and photoelectron yield measurements as well
as the operational HV settings for 12 modules. Additional data (current-draw
measurements, dark-count rates) are also collected using the test station.

The OPAL design features an optical port at the base of each 
module. Blue LEDs are installed in these optical ports
as part of the calibration and monitoring system, and will allow
monitoring of the operational status and relative timing for each 
block. A low-capacitance LED was chosen to minimize the rise and fall
times of the light pulse; this is important for use with the 
time-over-threshold-based readout system discussed in the following section.
In principle, the LED system should allow in-situ gain measurement as well. 

\begin{figure}
\centering
\includegraphics[width=0.725\linewidth]{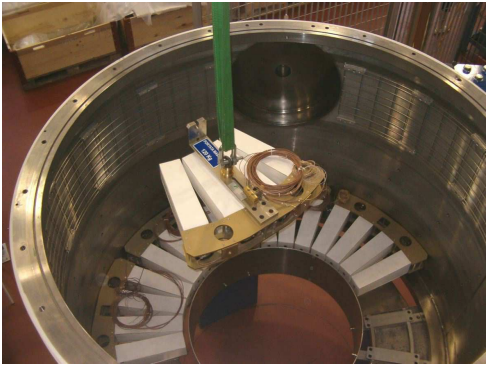}
\caption{Installation of lead glass modules into the steel vacuum vessel.}
\label{fig:assy}
\end{figure}
After testing and characterization, the blocks are arranged in groups of 
four in an aluminum mounting bracket. For the installation, the vacuum 
vessel is turned on end. The aluminum mounting bracket with four modules
is lowered into the upended vessel by overhead crane and bolted to the 
wall, as illustrated in \Fig{fig:assy}. The HV, signal, and LED cables are
then routed to the flanges along the cable grille visible at the top of the 
photo. In order to preserve uniformity of signal shapes and timing across
channels, all signal (and LED) cables have the same length, with cable slack
taken up on the grille. For HV cables, the excess length is cut.

As noted above, the signal, HV, and LED connections are made on five 200-mm 
ISO-F flanges. Flanges with eight or ten DB37 connectors are used both for 
signal and LED feed-through, 16 channels per connector, with separate ground
connections for each signal channel. Flanges with eight 
32-pin MIL-C-26482 connectors are used for HV feed-through, 32 channels per 
connector, with a common ground for all 32 channels on a separate MHV 
feedthrough.
  
\section{Front-End Electronics}

As noted above, the LAVs must furnish time and energy measurements over  
a large range of incident photon energies. For reasons of cost and simplicity,
we have decided on a readout scheme based on the time-over-threshold (ToT) 
technique. This scheme is implemented using a dedicated front-end ToT 
discriminator board of our own design \cite{A+11:ToT} and a digital
readout board (TEL62) used by various NA62 detector subsystems 
\cite{A+11:TEL62}.   
The ToT discriminator converts the analog signals from 
the detector to low-voltage differential signal (LVDS) pulses, with width 
equal to the duration of the analog signal from the detector above a 
specified threshold. The signal from each PMT is compared to two different
thresholds, a low threshold of about 5~mV and a high threshold expected to be
about 50~mV, corresponding to two different LVDS outputs. (For comparison, 
the expected noise level is 2~mV under normal operating conditions.) 
The TEL62 is based on the design of the TELL1 readout board developed for the 
LHCb experiment. On the TEL62, TDC mezzanines measure the leading and 
trailing times of the LVDS pulses. Using a time-to-charge calibration 
parameterization, the FPGA on board the TEL62 calculates the time 
corrected for slewing, as well the charge for each hit as reconstructed 
from the pulse width above threshold. This information is sent to the 
subsequent DAQ stages. Level-0 trigger primitives are also calculated on 
board the TEL62 and sent to the level-0 trigger processor.

\begin{figure}
\centering
\includegraphics[width=0.9\linewidth]{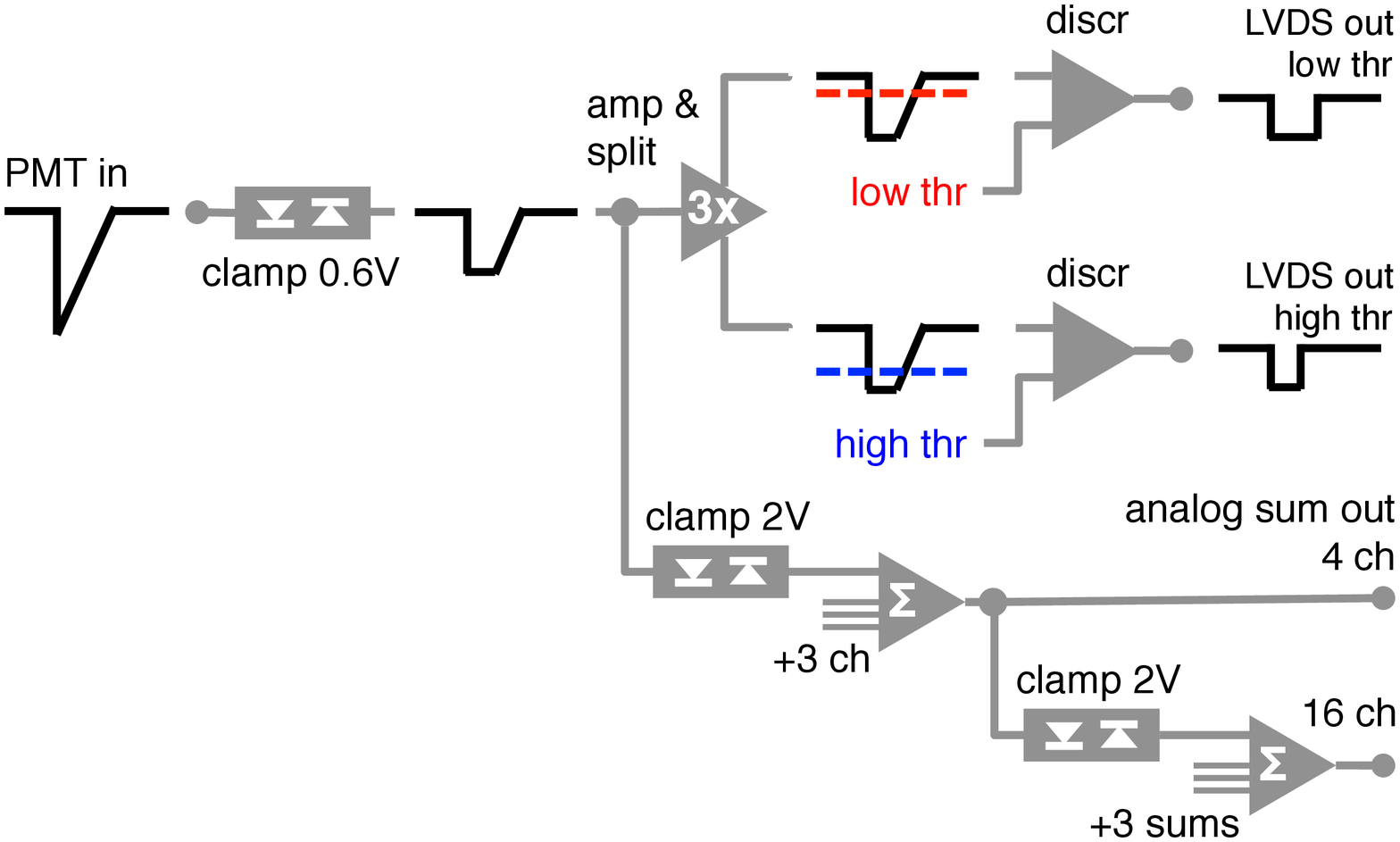}
\caption{Conceptual schematic of the ToT discriminator board.}
\label{fig:fee}
\end{figure}
A conceptual schematic of the ToT discriminator board is presented in 
\Fig{fig:fee} to illustrate the signal processing for a single channel.
Since each channel is discriminated against two different
thresholds, the board has 32 input channels and 64 output channels.

While the amplitude of the PMT signal from a minimum ionizing 
particle is about 20~mV, signals from 20-GeV showers may be as large as 10~V.
For protection, the input signal is clamped at 0.6~V by a circuit that 
maintains the timing of the rising and falling edges of the pulse. The
clamped signal is then passively split. One copy of the signal is summed 
with the signals from other channels to form diagnostic analog outputs
as discussed below. The other copy is amplified $\times3$ by a low-noise,
high-bandwidth, high-speed amplifier and passively split into two copies.
Each copy is used as input to a high-speed comparator with an LVDS driver.
The threshold for comparison is provided by a programmable DAC in the 
board controller. To reduce double pulses from the comparator due to noise
in the input signal, 3 mV of hysteresis is also provided through a feedback 
resistor, so that the output signal is extended until the input signal falls
to 3 mV below the threshold. 

The other copy of the clamped analog signal is summed with the signals from 
three adjacent channels (and clamped at 2~V) for diagnostic purposes. 
The resulting analog sum is made available via a front-panel LEMO connector.
These sums of four are in turn summed four at a time to produce sums of 16.
For 32 input channels, there are a total of 10 front-panel analog outputs:
eight sums of four and two sums of 16. A compact way to perform single channel 
diagnostics is to pulse the blocks one at a time using the LED system and 
read out the analog signal from these sums. 

The readout board is implemented on a 9U VME card and has a modular structure,
with many of the important functions on mezzanines, to reduce costs and 
simplify repairs. The ToT discriminators 
are implemented on 16 mezzanines, each with 4 discriminator circuits.
The analog sums are handled on 10 mezzanines (one per front-panel output).
A pulse-generator mezzanine allows pulses of programmable width and amplitude 
to be sent to a pattern of channels on the input connectors.
The board controller mezzanine handles communication with the experiment's
slow-control system to allow the thresholds to be set and read, to
control the test-pulse generator, and to monitor the board status.

\section{Test-Beam Performance}

The A2 LAV station was tested using the positive secondary beam in the T9
area at the CERN PS in August 2010. Data were collected at various beam
momenta over the interval 0.3--10~GeV. 
The composition of the T9 beam changes as a function of the momentum setting.
At 0.3~GeV, the beam is roughly 70\% $e^+$ and 30\% $\pi^+$ (including 
decay $\mu^+$), while above
6~GeV the $e^+$ component drops off sharply and at 10~GeV the beam is 
roughly a 50\% mix of $\pi^+/\mu^+$ and $p$.
Two threshold Cerenkov counters in the beamline using CO$_2$ at adjustable 
pressure allowed samples enriched $e^+$ and $\mu^+$ to be selected. 
The T9 detectors also included two scintillation paddles and a beam wire 
chamber for beam counting and diagnostics. The beam focus was 2~m 
downstream of the beam wire chamber, and A2 was positioned so that the 
beam focus coincided approximately with the first layer of lead glass 
blocks. Two additional scintillator paddles ($60\times85$~mm$^2$) were placed
in a cross configuration at beam entrance and one larger paddle was placed
at beam exit. Requirements on the signal amplitude for both entrance counters 
significantly enriched the sample in events with a single beam particle 
incident on a single column of lead-glass blocks, while requiring in 
addition a hit on the exit counter significantly enriched the sample 
in events in which the beam particle was a $\mu^+$. 
A prototype of the ToT discriminator board (with one discriminator per 
channel, rather than two) was tested 
together with A2. For much of the test, our board was used together with 
commercial QDCs and TDCs, but in dedicated runs, the times were also 
measured with a prototype setup using a TELL1 board from LHCb 
and an early version of the TDC mezzanine.

\begin{figure}
\centering
\includegraphics[width=0.85\linewidth]{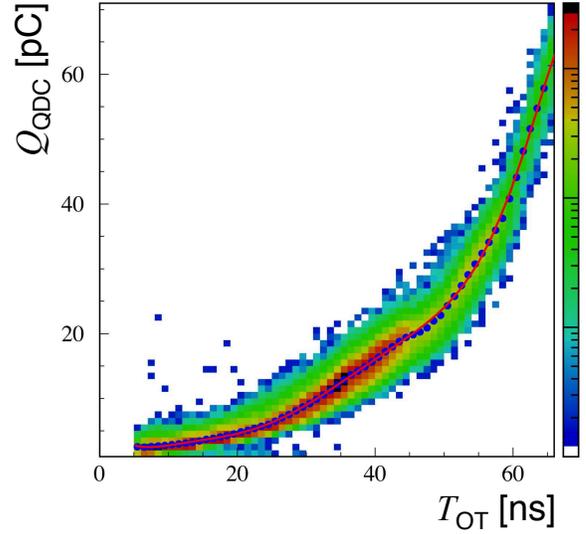}
\caption{Scatter plot of signal charge measured using the QDCs vs.\ signal 
time over threshold, for electrons of various energies at the 
T9 test beam.}
\label{fig:qtot}
\end{figure}
\begin{figure}
\centering
\includegraphics[width=0.85\linewidth]{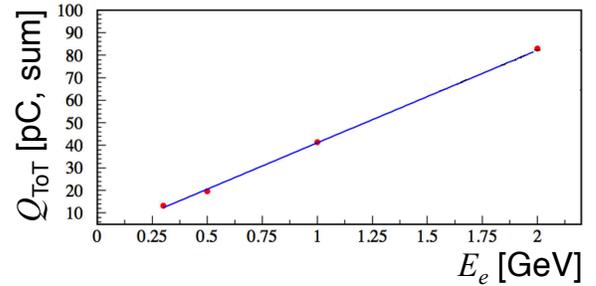}
\caption{Equivalent charge-integrated signal from ToT vs.\ incident 
electron energy, demonstrating linearity of response.}
\label{fig:lin}
\end{figure}
\begin{figure}
\centering
\includegraphics[width=0.85\linewidth]{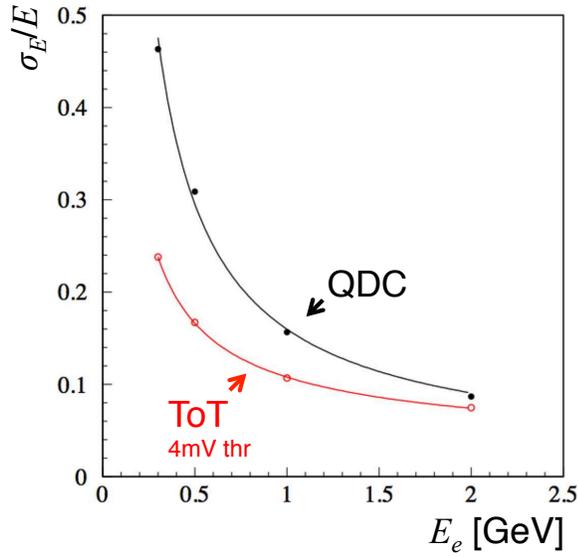}
\caption{Energy resolution obtained using the ToT technique compared with 
that obtained using the QDCs, as a function of incident electron energy.}
\label{fig:eres}
\end{figure}
Figure~\ref{fig:qtot} shows a scatter plot of the signal charge 
measured using the QDCs vs.\ the signal time over a threshold of 4~mV, for 
electrons. The data are summed for all blocks on which the beam was 
incident and over runs of different energies. For small signals, a
small increase in the integrated charge corresponds to a large increase 
in the time over threshold; the ToT measurement 
provides more sensitivity for small signals than does the QDC measurement.
This is a desirable property for the LAV detectors, since high detection 
efficiency is required for low-energy photons.
Parameterizations of the type illustrated in \Fig{fig:qtot}
are used to convert the time over threshold to an effective charge 
measurement.
Figure~\ref{fig:lin} demonstrates that good linearity of response
is obtained using this method. The plot shows the equivalent charge in 
pC from the ToT measurement (with 4~mV threshold) for electrons of
energy 0.3, 0.5, 1, and 2~GeV. Linearity to within 1\% is observed.
The energy resolution obtained using the ToT technique is compared with that 
obtained using the QDCs in \Fig{fig:eres}. As expected from the form of 
the curve in \Fig{fig:qtot}, at low energies,
the resolution obtained with the ToT technique is better than that obtained 
with the QDCs. The fits in \Fig{fig:eres} give
\begin{eqnarray*}
{\rm QDC:}&\hspace{1mm}& \sigma_E/E = 8.6\%/\sqrt{E~{\rm[GeV]}} \oplus 13\%/E, \\
{\rm ToT:}&\hspace{1mm}& \sigma_E/E = 9.2\%/\sqrt{E~{\rm[GeV]}} \oplus 5\%/E \oplus 2.5\%.
\end{eqnarray*}
While, as expected, the statistical contribution to the energy resolution 
is about the same with either readout scheme, the contribution from noise 
(term proportional to $1/E$) appears to be significantly smaller with the ToT 
technique. The presence of the constant term with the ToT technique may be 
due to small differences in the charge vs.\ ToT curves from block to block, 
and if so, can potentially be reduced.

\begin{figure}
\centering
\includegraphics[width=0.85\linewidth]{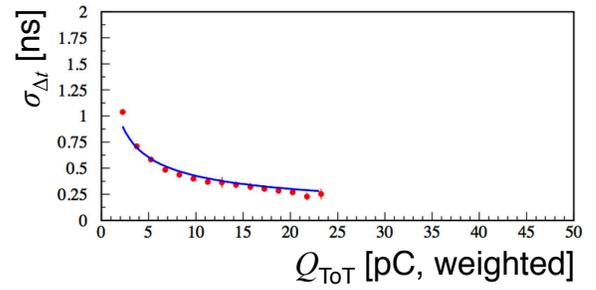}
\caption{Time resolution obtained using the ToT technique
as a function of incident electron energy.}
\label{fig:tres}
\end{figure}
To study the time resolution, it is first necessary to correct for slewing.
The event time reference is obtained from the scintillator paddles and 
Cerenkov counters, and the curve of signal time vs.\ integrated charge from 
ToT is parameterized assuming that the signal shape is well described by the 
form $V(t) \propto t^a e^{-bt}$. The width of the signal time distribution 
in slices of charge gives a measurement of the time resolution. A fit to the 
measurements of $\sigma_t$ vs. charge from ToT (4~mV threshold) for a single 
block gives $\sigma_t = 220~{\rm ps}/\sqrt{E~{\rm[GeV]}} \oplus 140~{\rm ps}$, 
where the constant term is assumed to be due to trigger jitter. This 
assumption can be tested by measuring the width of the distribution of 
signal time differences for two successive blocks.
Assuming that the two blocks have the same intrinsic time response and that 
there is no common-mode contribution to the resolution,
$\sigma_{t_1} = p/\sqrt{E_1}$ and $\sigma_{t_2} = p/\sqrt{E_2}$,
so that we expect $\sigma_{\Delta t} = p/[E_1 E_2/(E_1 + E_2)]^{1/2}$, with no
constant term. Figure~\ref{fig:tres} shows the measurements of 
$\sigma_{\Delta_t}$ in slices of the weighted charge measurements from ToT,  
$Q = Q_1Q_2/(Q_1 + Q_2)$ (4~mV threshold). The overlaid curve has $p = 210$~ps
and no constant term, which is indicative of the intrinsic time resolution
of the detector.   

\begin{figure}
\centering
\includegraphics[width=0.85\linewidth]{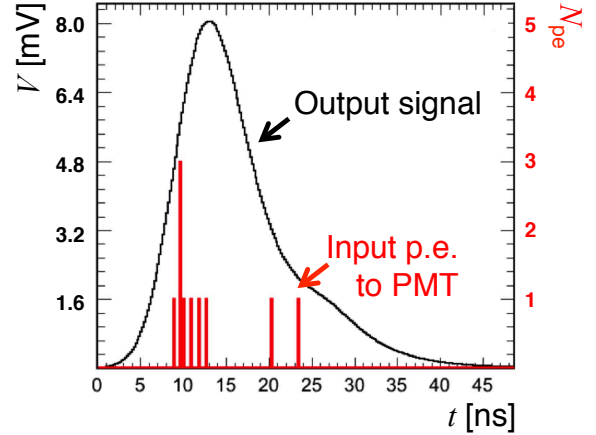}
\caption{Simulation of PMT signals. The red histogram shows the number of 
photoelectrons produced in bins of time (scale at right). The black curve is
the resulting PMT signal (scale at left).}
\label{fig:sig}
\end{figure}
\begin{figure*}
\centering
\includegraphics[width=0.75\linewidth]{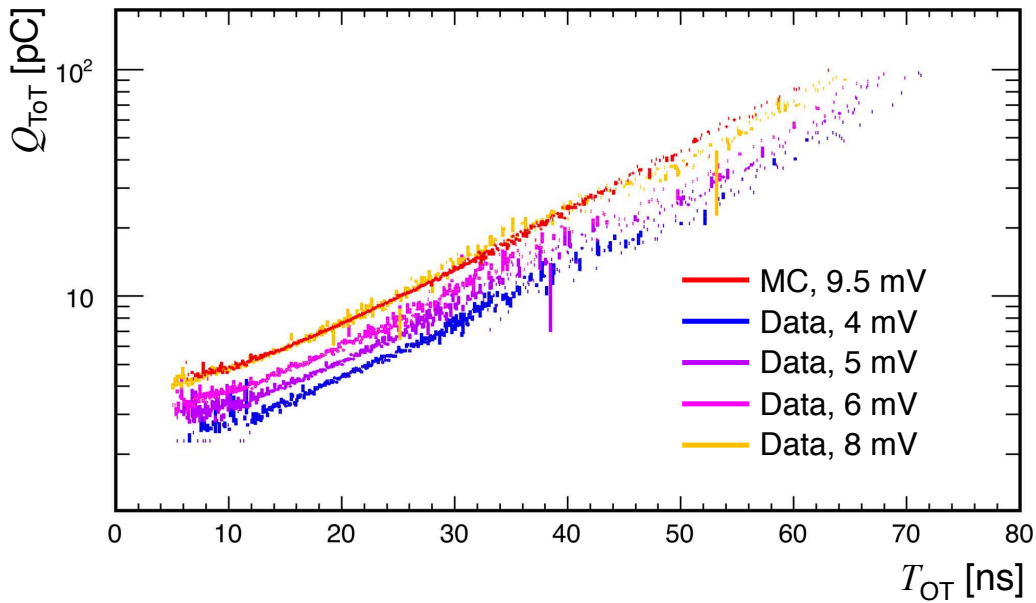}
\caption{Comparison of charge vs.\ ToT curves for MC (9.5 mV threshold) and 
data (various thresholds). Apart from a small discrepancy in the value of the 
threshold, the simulation accurately reproduces the shape of the charge vs.\ 
ToT curve.}
\label{fig:sim}
\end{figure*}
The results obtained at the beam test provide a point of comparison for
the Monte Carlo (MC) simulation of the LAV system. The Geant4-based NA62 
MC includes a detailed description of the LAV geometry and materials.
Two sets of routines are available for shower simulation and the creation 
and propagation of Cerenkov photons: the standard Geant4 routines, with 
complete tracking of the optical photons, and our own routines, which make
use of a response matrix obtained from the full simulation. In either case,
the simulation produces the number of Cerenkov photons that arrive at the 
photocathode, together with their arrival times, as shown by the 
histogram in red in \Fig{fig:sig}. A complete simulation of
the PMT uses this information to generate an output signal, taking into 
account the PMT gain and transit-time fluctuations, the capacitance of the 
PMT, and dispersion in the readout cable. The resulting signal, which
is illustrated by the black curve in \Fig{fig:sig}, is input to 
a full simulation of the ToT discriminator, including the comparator 
hysteresis. The complete simulation thus yields a description of the PMT 
signal, the integrated charge, and the ToT response.

Figure~\ref{fig:sim} shows charge vs.\ ToT curves for test-beam muons for
four different threshold settings (4, 5, 6, and 8~mV). The red curve is 
from the simulation, with a threshold at 9.5~mV and all other adjustable 
parameters (e.g., tube gain, interdynode time fluctuations, PMT capacitance) 
set to typical or measured values. The simulation with a threshold of 9.5~mV
accurately reproduces the shape of the measured charge vs.\ ToT curve with
threshold 8~mV. Apart from this slight discrepancy, which is easily attributed 
to the accuracy of the manual threshold adjustment on the prototype ToT 
discriminator board, these results confirm our detailed understanding of the
detector and the readout chain. 

\section{Status and Outlook}

\begin{figure}
\centering
\includegraphics[width=0.8\linewidth]{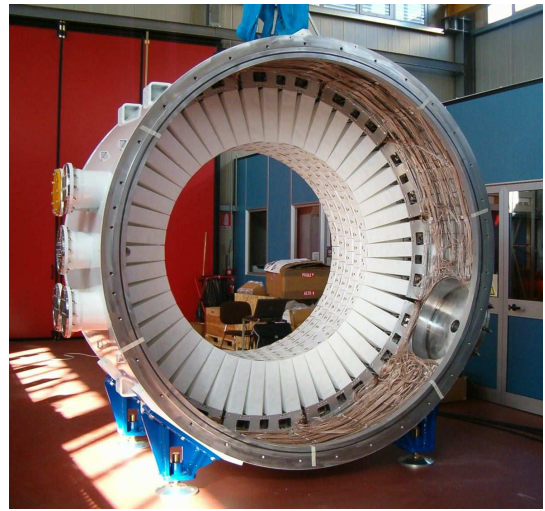}
\caption{The first intermediate diameter LAV station (A6) upon completion
at Frascati in November 2011.}
\label{fig:A6}
\end{figure}
Construction of the five small-diameter stations (including updating of 
the A1 prototype) was completed with all stations delivered to CERN and 
awaiting installation on the beamline in July 2011. 
Installation of these five LAVs is scheduled for December 2011. 
As of November 2011, construction of the remaining 
stations is in progress; the first of the
intermediate diameter stations was recently completed (\Fig{fig:A6}).
Serial production of the front-end electronics boards is beginning. 
The installed detectors will be read out in a dry run in mid 2012, while 
a technical run with beam is scheduled for late 2012 and will include at 
least eight of the twelve LAVs. Data taking with the remaining detectors 
is planned for 2013.

\section*{Acknowledgments}

We warmly thank C. Capoccia and A. Cecchetti of the Experimental Apparatus 
Design Service (SPAS) at the INFN Frascati Laboratories, and S. Bianucci 
of INFN Pisa, for their collaboration on the mechanical design of the LAV
system, including the steel vacuum vessels. We also thank our mechanical 
design team for their assistance in oversight of the construction of the 
steel vessels and of the transport of the completed detectors to CERN.
We thank G. Corradi, D. Tagnani, and C. Paglia of the Electronics Service 
(SELF) at INFN Frascati for their collaboration on the design of the new 
HV dividers and front-end electronics boards.
Construction was made possible by the contributions from our INFN 
technicians: E. Capitolo, R. Lenci, V. Russo, M. Santoni, T. Vassilieva, 
and S. Valeri (Frascati); F. Cassese and L. Roscilli (Napoli); 
L. Berretta and G. Petragnani (Pisa); and F. Pellegrino (Roma).
Finally, We thank V. Lollo and P. Chimenti of the Vacuum Service at 
INFN Frascati for their assistance with vacuum-related issues. 

\bibliographystyle{IEEEtran}
\bibliography{valencia_proc}

\end{document}